\documentclass[journal]{IEEEtran}

\usepackage{graphicx}
\usepackage{setspace}
\usepackage{acronym}
\usepackage{amssymb}
\usepackage[cmex10]{amsmath}
\usepackage{commath}
\usepackage{mathrsfs}
\usepackage{mathtools}
\usepackage{ifthen}
\usepackage{textcomp}
\usepackage{float}
\usepackage[tight,footnotesize]{subfigure}
\usepackage{cite}
\usepackage{bm}
\usepackage{url}

\usepackage{siunitx}

\usepackage{multirow}
\usepackage{array}
\newcolumntype{L}[1]{>{\raggedright\let\newline\\\arraybackslash\hspace{0pt}}m{#1}}
\newcolumntype{C}[1]{>{\centering\let\newline\\\arraybackslash\hspace{0pt}}m{#1}}
\newcolumntype{R}[1]{>{\raggedleft\let\newline\\\arraybackslash\hspace{0pt}}m{#1}}

\IEEEoverridecommandlockouts

\begin{document}

\title{SMIET: Simultaneous Molecular Information and Energy Transfer}

\author{\authorblockN{Weisi Guo\textsuperscript{1}, Yansha Deng\textsuperscript{2}, H. Birkan Yilmaz\textsuperscript{3}, Nariman Farsad\textsuperscript{4}, Maged Elkashlan\textsuperscript{5},\\ 
Chan-Byoung Chae\textsuperscript{3}, Andrew Eckford\textsuperscript{6}, and Arumugam Nallanathan\textsuperscript{2}}

\thanks{\textsuperscript{1}W. Guo is with the University of Warwick, UK. \textsuperscript{2}Y. Deng and A. Nallanathan are with King's College London, UK. \textsuperscript{3}H. B. Yilmaz and C.-B. Chae are with Yonsei University, Korea. \textsuperscript{4}N. Farsad is with Stanford University, USA. \textsuperscript{5}M. Elkashlan is with Queen Mary University of London. \textsuperscript{6}A. Eckford is with York University, Canada. Corresponding Author: cbchae@yonsei.ac.kr}}

\maketitle

\begin{abstract}
The performance of communication systems is fundamentally limited by the loss of energy through propagation and circuit inefficiencies. In this article, we show that it is possible to achieve ultra low energy communications at the nanoscale, if diffusive molecules are used for carrying data. While the energy of electromagnetic waves will inevitably decays as a function of transmission distance and time, the energy in individual molecules does not. Over time, the receiver has an opportunity to recover some, if not all of the molecular energy transmitted. The article demonstrates the potential of ultra-low energy simultaneous molecular information and energy transfer (SMIET) through the design of two different nano-relay systems. It also discusses how molecular communications can benefit more from crowd energy harvesting than traditional wave-based systems. 
\end{abstract}


\section{Introduction}

Over the past decade, there has been a growing focus on increasing the energy efficiency of both mobile and fixed wireless systems. While we have built up a good understanding of power consumption mechanisms in terrestrial mobile networks, we still lack understanding in how nano-machines\footnote{Nano-machines are devices that are constructed using nanoscale technology and typically have dimensions of 1-10 microns.} can communicate in an energy-efficient manner. It is envisaged that nano-scale communications will be critical to nano-machines that seek to coordinate tasks such as in vivo drug delivery and surgery \cite{Akyildiz15}. While many meso- and macro-scale in vivo medical devices (i.e., pacemaker) are battery powered, nano-batteries (50 microns \cite{Ajayan11Nano}) are still significantly larger than or of the same dimension as their nano-machine counter-parts. Therefore, charging batteries using externally generated acoustic \cite{Donohoe16} and electromagnetic radiation is not always viable and furthermore, nano-machines can be embedded in vivo areas that are either sensitive to radiation or difficult for radiation to penetrate. What is needed in such cases is energy harvesting from the nano-machines' locality. Over the past decade, more and more studies have explored ways to harvest energy from communication signals and achieve simultaneous wireless information and power transfer (SWIPT). In this article, we draw on our understanding of energy consumption and on our knowledge of harvesting in current wireless systems to better understand nano-scale communications and exploit opportunities. In so doing, we propose simultaneous molecular information and energy transfer (SMIET). 

\subsection{Nano-Scale and Molecular Communications}

The traditional practices of wireless planning with known coverage areas and propagation environments starts to breakdown at the micro- and nano-scales. Communication systems in complex biological environments must have the following characteristics: bio-compatibility, low power consumption, low complexity, small dimension, and the ability to achieve reliable signaling in a fluid environment with complex cell/tissue obstacles. Such constraints are challenging for both EM-based THz systems and nano-acoustic systems. Inspired by the abundant use of molecules in biological signaling, Molecular Communication via Diffusion (MCvD) utilizes \emph{molecular signal} (i.e., a chemical pulse) as an alternative carrier of information \cite{Farsad16_Survey}. MCvD avoids the limitations of wave generation and propagation, and allows the signal to both persist and propagate to areas that are difficult to reach. Indeed, this is perhaps why MCvD is prevalent in nature, both at the inter-organism and at the inter-cell scales. In terms of application, the capacity of nano-scale devices to communicate with one another may have unlocked the potential of nano-robotics in medicine \cite{Akyildiz15}. The economic potential is enormous: the present market size is at \$100 billion, growing at a projected 14\% per year (BCC research). 
This article speculates on the potential of ultra-low energy communications using molecular message carriers, especially in the context of nano-communications. Nano-machines are likely to require low data rate communications that lack the energy and complexity to perform coherent channel estimation, synchronization, and range or location estimation. Therefore, nano-machines need to perform basic mechanical tasks and communicate to each other while expending a low amount of energy. We highlight the potential of energy harvesting using molecular communications and we also design efficient and reliable systems.

\subsection{Biological Energy Harvesting Systems}

\begin{figure}[tb]
	\centering
	\includegraphics[width=1.00\linewidth]{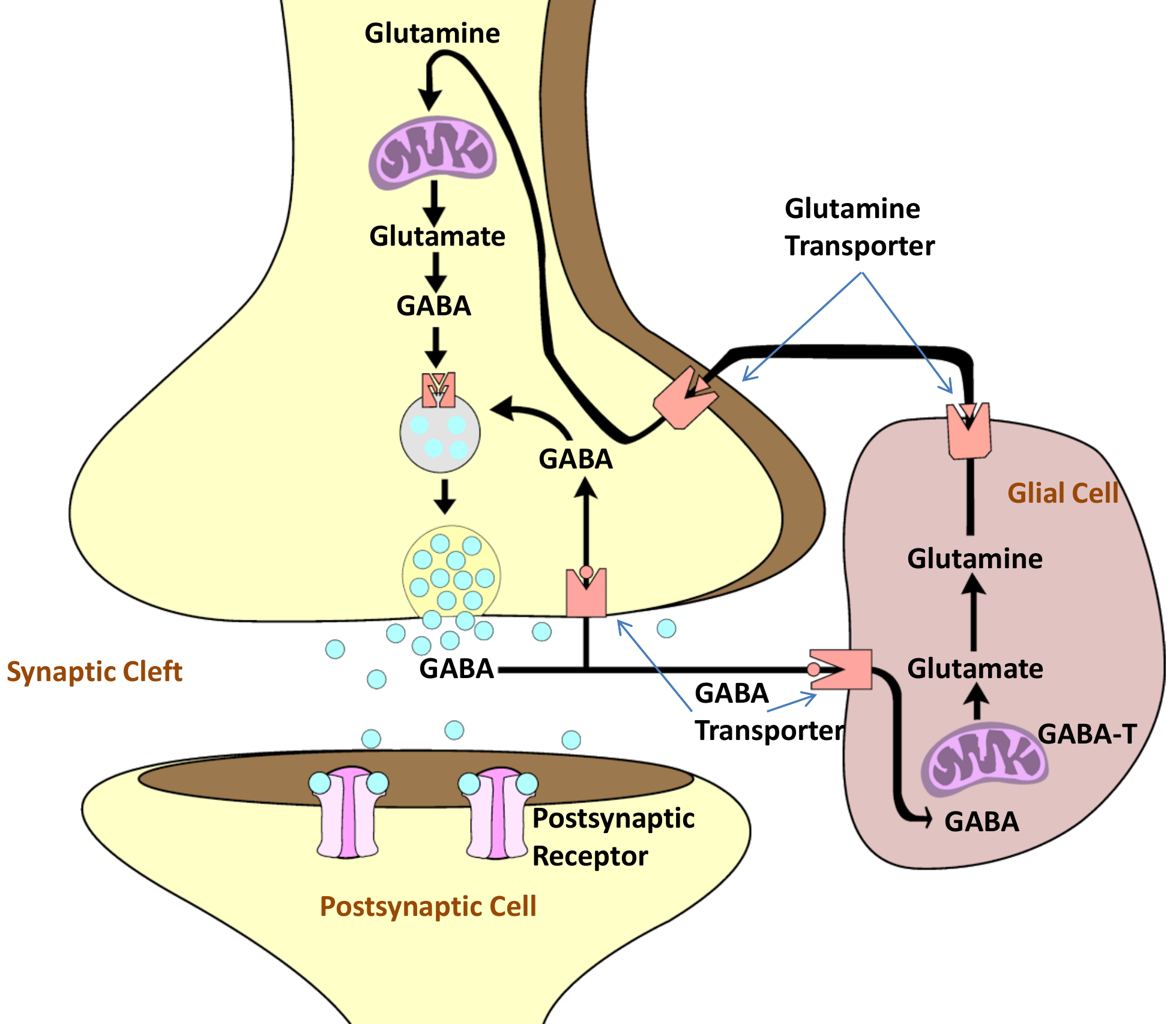}
	\caption{GABA reuptake mechanism at the synaptic cleft. Glial cell absorbs/harvests GABA molecules and converts them to Glutamine for utilizing at the signaling mechanisms of presynaptic region.}
	\label{GABA}
\end{figure}

In fact, in biology at the cellular-level it is common to harvest energy using signaling molecules. One such biological example that utilizes two types of molecules is depicted in Fig.~\ref{GABA}. Type A molecule generated by the source is absorbed and converted to type B molecule inside the receiving node to be utilized in another subsequent signaling mechanism. 

The $\gamma$-Aminobutyric acid (GABA) metabolism and uptake is widely distributed across almost all regions of the mammalian brain. GABA is constructed by glutamate via enzymatic reaction with glutamic acid decarboxylase (GAD) in the presynaptic neuron cell, which is then released as a neurotransmitter for sending a signal to both neighbor Glia cells and postsynaptic neuron cells via GABA transporters (GATs). In this example, the presynaptic neuron cell acts as the source for emitting the GABA as a type A molecule. The Glia cell acts as the molecule harvesting node, and emitting glutamine as a type B molecule in response to the electrical charge polarization caused by GABA. 

This example of molecule harvesting and manipulation will be reverse engineered in this article to demonstrate how energy harvesting in MCvD is in some ways similar to current systems, but with new opportunities for exploitation. This article is organized as follows. In Section II, we discuss the general power consumption model for the generation, propagation, and reception of data. We compare MCvD and RF power consumption models and offer a physical explanation on why MCvD avoids the heavy propagation losses associated with wave-based communications. In Section III, we demonstrate how to achieve ultra-low energy communications via two nano-relay systems: (1) single molecule type with interference shielding, and (2) multiple molecule types with fragmentation and synthesis. In Section IV, we discuss how recent advances in relay-based Simultaneous Wireless Power Transfer (SWIPT) and crowd energy harvesting can be leveraged for MCvD. In Section V, we conclude the findings and discuss multi-disciplinary communication research opportunities going forwards.

\section{Power Consumption Model}

The power consumption model of a generic wireless system can be approximately modularized into a number of contributing components. In this article, we focus on the radio layer of consumption (including the data modulation, amplification, antenna, and propagation effects). Left to future discussions is the overhead consumption incurred by the signal processing and cooling elements. Such consumption consists chiefly of the following: i) the receiver antenna gain with radius $R$, ii) the free-space propagation loss $\lambda$, iii) the transmitter efficiency (i.e., power amplifier efficiency $\mu$ or chemical synthesis cost $\phi$), iv) absorption loss $\tau$ (also known as transmittance), and the v) the circuit power consumption. In general, the received electromagnetic (EM) power ($P_{\text{Rx}}$) or molecular number ($N_{\text{Rx}}$) is a small fraction of the total power extracted by the transmitter $P_{\text{Total}}$:
\begin{equation}\begin{split}
\label{Power_Equation}
\frac{P_{\text{Rx}}}{P_{\text{Total}}} &\propto \mu \times \tau \bigg(\frac{R^{2}}{d^{\alpha}}\bigg) \quad \mbox{for EM} \\
\frac{N_{\text{Rx}}}{P_{\text{Total}}} &\propto \frac{1}{\phi(n_{\text{Tx}}-1)} \times \bigg(\frac{R}{d+R}\bigg) \quad \mbox{for MCvD: } t\to \infty,                                         
\end{split}\end{equation} assuming that the communication circuit power is relatively small in a nano-machine. Like RF communications, there is an energy efficiency factor for generating $N_{\text{Tx}}$ molecules for transmission. This is related to the number of basic chemical components per molecule $n_{\text{Tx}}$ and the energy cost to bind or synthesize them $\phi$ \cite{Furubayashi16, Kuran10}. In the rest of the section, we now explain the reasoning and details of the efficiency equations given in \eqref{Power_Equation}.

\subsection{Electromagnetic (EM) Wave Communications}
\begin{figure}[t]
	\centering
	\includegraphics[width=1.00\columnwidth]{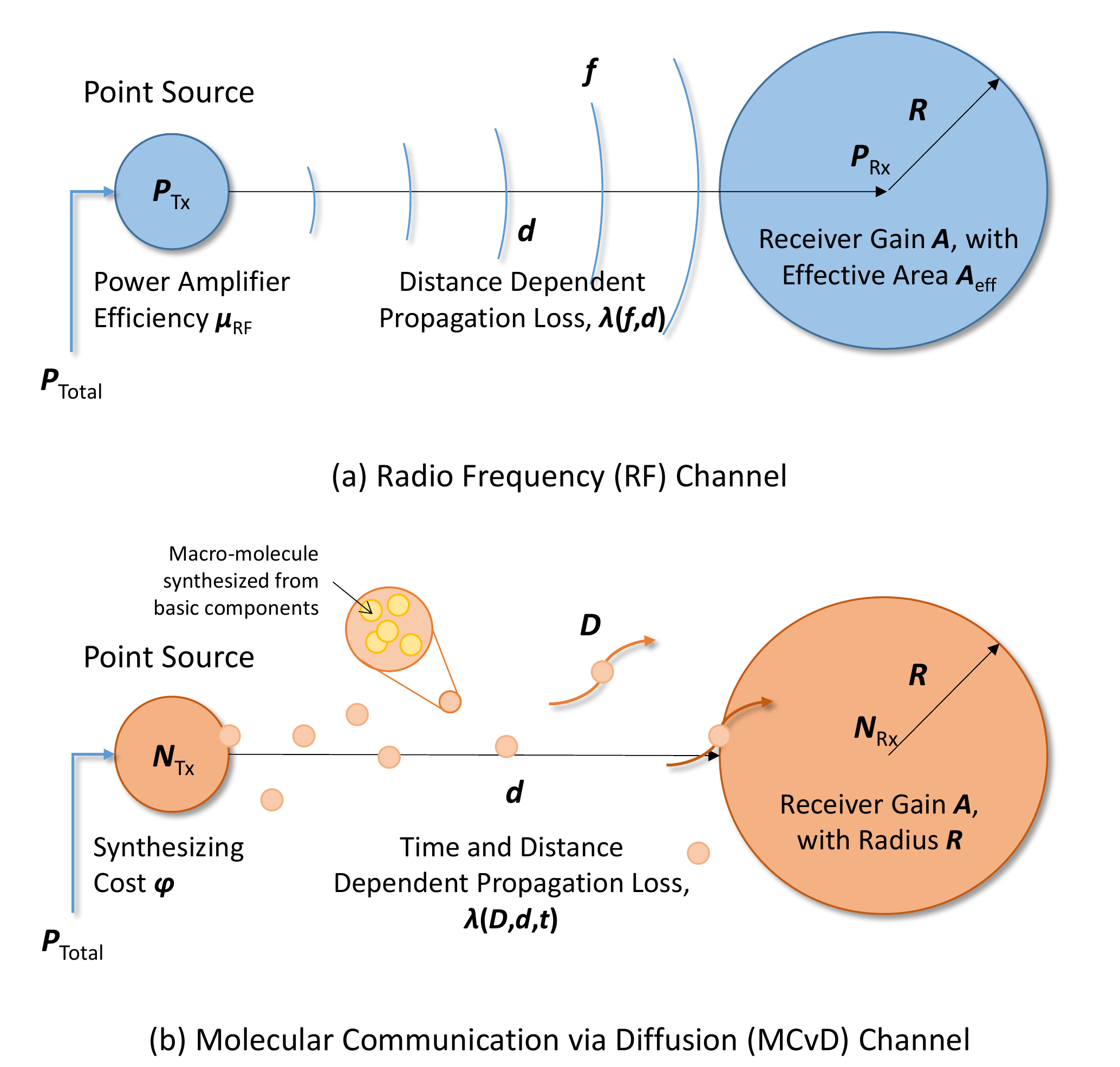}
	\caption{Illustration of power loss in transmitting signals in (a) Electromagnetic (EM) Wave communications, and (b) Molecular Communications via Diffusion (MCvD).}
	\label{Energy_Model}
\end{figure}

We first examine the EM communications case. One can see that the best achievable efficiency in~\eqref{Power_Equation} is limited by $d^{-\alpha}$ for RF and limited by $\tau d^{-\alpha}$ for THz nano-scale communications (where $\tau \propto \exp(-kfd)$). The absorption loss $\tau$ is log-linear proportional to absorption coefficient $k$, which depends on the chemical composition of the medium and is typically $1\times10^{-5}$ for air and $1-3$ for water at $f=0-\SI{10}{ {THz}}$~\cite{Akyildiz11}. To demonstrate the aforementioned reasoning, we show an illustration of radio wave versus molecular propagation in Fig.~\ref{Energy_Model}. In the (a) subplot, an isotropic EM antenna transmits to a receiver antenna with an effective area $A_{\text{eff}} \propto (fR)^{2}$, and the resulting received power after propagating a distance of $d$ is $\mu_{\text{EM}}\frac{R^{2}}{d^{-\alpha}}$, where the value of $\mu_{\text{EM}}$ is typically 10-30\% for RF systems \cite{Auer11}.

\subsection{Molecular Communication via Diffusion (MCvD)}
MCvD on the other hand, relies on message bearing molecules to be freely diffused from the transmitter to the receiver. We do not consider the base energy cost of physical matter (i.e., the molecules) as matter is not lost in the communication process. We do, though, consider the energy cost of creating specific chemical compounds, as well as the energy benefits of restructuring the compound. 

In general, MCvD involves messenger molecules performing a random-walk motion across the communication channel through collision interaction and a diffusion gradient. For each emitted molecule, there is a finite probability that it will reach the intended receiver. The power in the system at any given time instance is proportional to the number of molecules. While the stochastic process is intuitively unreliable and requires a transmission time that is orders of magnitude longer than wave propagation, these deficiencies can be mitigated by communicating at very small distances (microns) or with the aid of a strong ambient flow. For a basic random-walk process, consider (as in Fig.~\ref{Energy_Model}b) a point emitter that transmits $N_{\text{Tx}}$ molecules. The full absorption receiver will capture $N_{\text{Rx}}$ molecules given by \cite{Yilmaz14}:
\begin{equation}\begin{split}
    \label{Capture}
    N_{\text{Rx}} &= N_{\text{Tx}}h_{c}, \quad h_{c} =\! \bigg(\frac{R}{d+R}\bigg)\frac{d}{\sqrt{4\pi D t^{3}}} \,e^{ \frac{-d^{2}}{4Dt} },
\end{split}\end{equation} where $h_{c}$ is known as the \textit{first passage time} density distribution.

The resulting expected number of received molecules up to time $t=T$ is $N_{\text{Tx}}F_{c}$, where $F_{c} = \frac{R}{d+R} \text{erfc}(\frac{d}{\sqrt{4DT}})$. This converges to $N_{\text{Tx}}$ for 1-dimensional (${\mbox{1-D}}$) space and $N_{\text{Tx}}\frac{R}{d+R}$ for ${\mbox{3-D}}$ space as $t \!\to\! \infty$. This means the full harvest of all transmitted molecules is possible in certain conditions, independent of the transmission distance. Naturally, the reality is that molecules will have a half-life and not all data-bearing molecules can be harvested. Reactions with other chemicals (i.e., enzymes) in the environment can, over time, reduce the effectiveness of energy harvesting in MCvD \cite{Noel14Enzyme}. Yet, the potential to capture the vast majority of the transmitted molecules due to the random-walk nature of propagation demonstrates the potential of MCvD over wave-based transmission. As with RF communications, there is a cost to producing the $N_{\text{Tx}}$ molecules at the source in the first place. This can be shown to be \cite{Kuran10}: $P_{\text{Total}} = \phi(n_{\text{Tx}}-1)N_{\text{Tx}}$, where $\phi$ is the synthesizing cost of bonding $n_{\text{Tx}}$ amino acids per transmitted molecule. 
\begin{figure}[t]
	\centering
	\includegraphics[width=1.00\columnwidth]{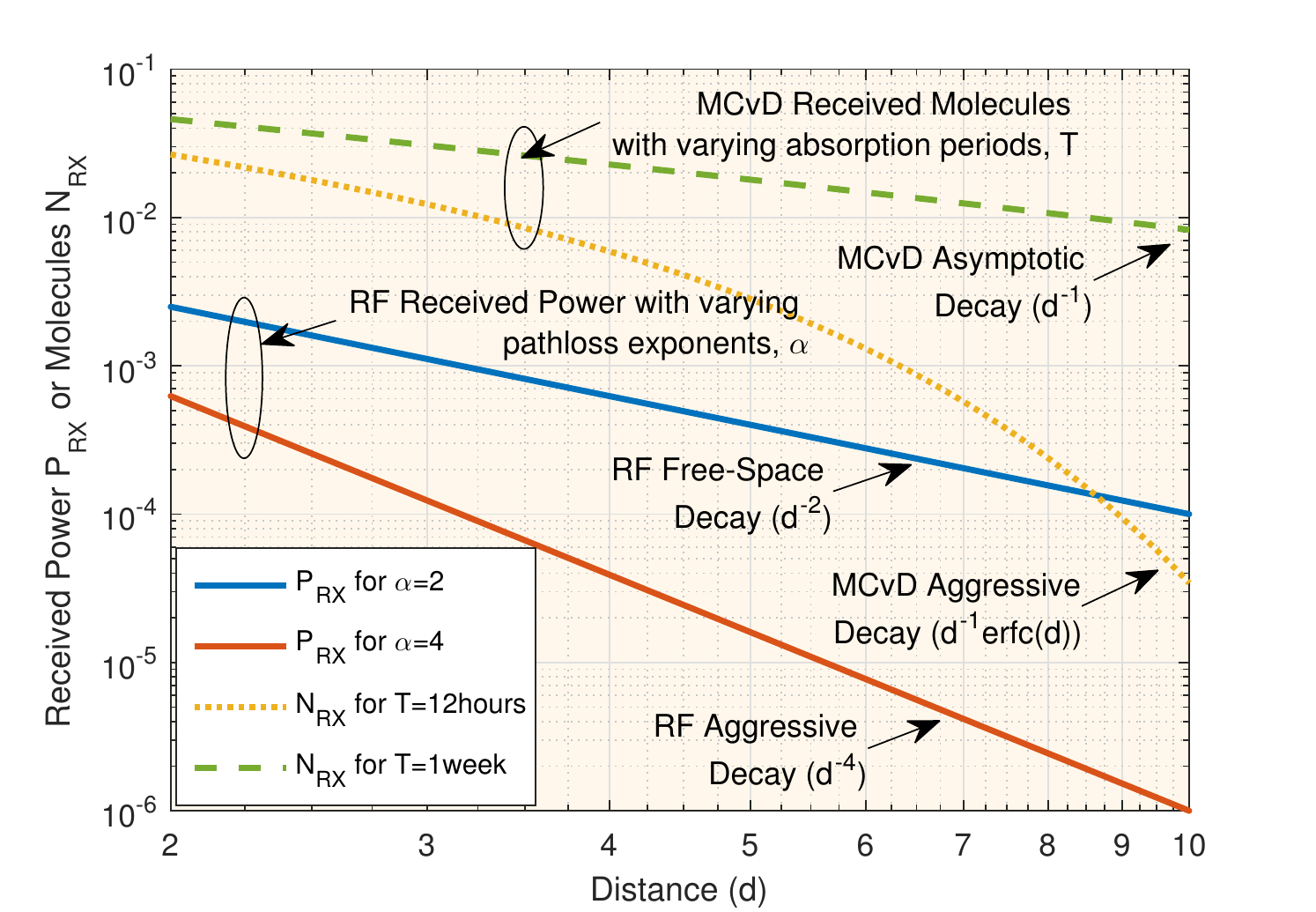} 
	\caption{Plot of received EM RF power ($P_{\text{Rx}}$) or MCvD molecules ($N_{\text{Rx}}$) for different different transmission distances ($d$). The results show that MCvD can achieve asymptotic distance-dependent power decay at the rate of $\propto d^{-1}$, which is superior to all RF scenarios. Modeling parameters: mass diffusivity (water molecules in air) $D=\SI{0.28}{\centi\metre^2/\second}$, RF frequency $\SI{5}{\giga\hertz}$ with parabolic receiver antenna $A_{\text{eff}} = 0.56{\pi R^{2}}$, and receiver radius of $R=\SI{10}{\centi\metre}$.}
	\label{Distance_Decay}
\end{figure}
\begin{figure*}[t]
	\centering
	\includegraphics[width=1.8\columnwidth]{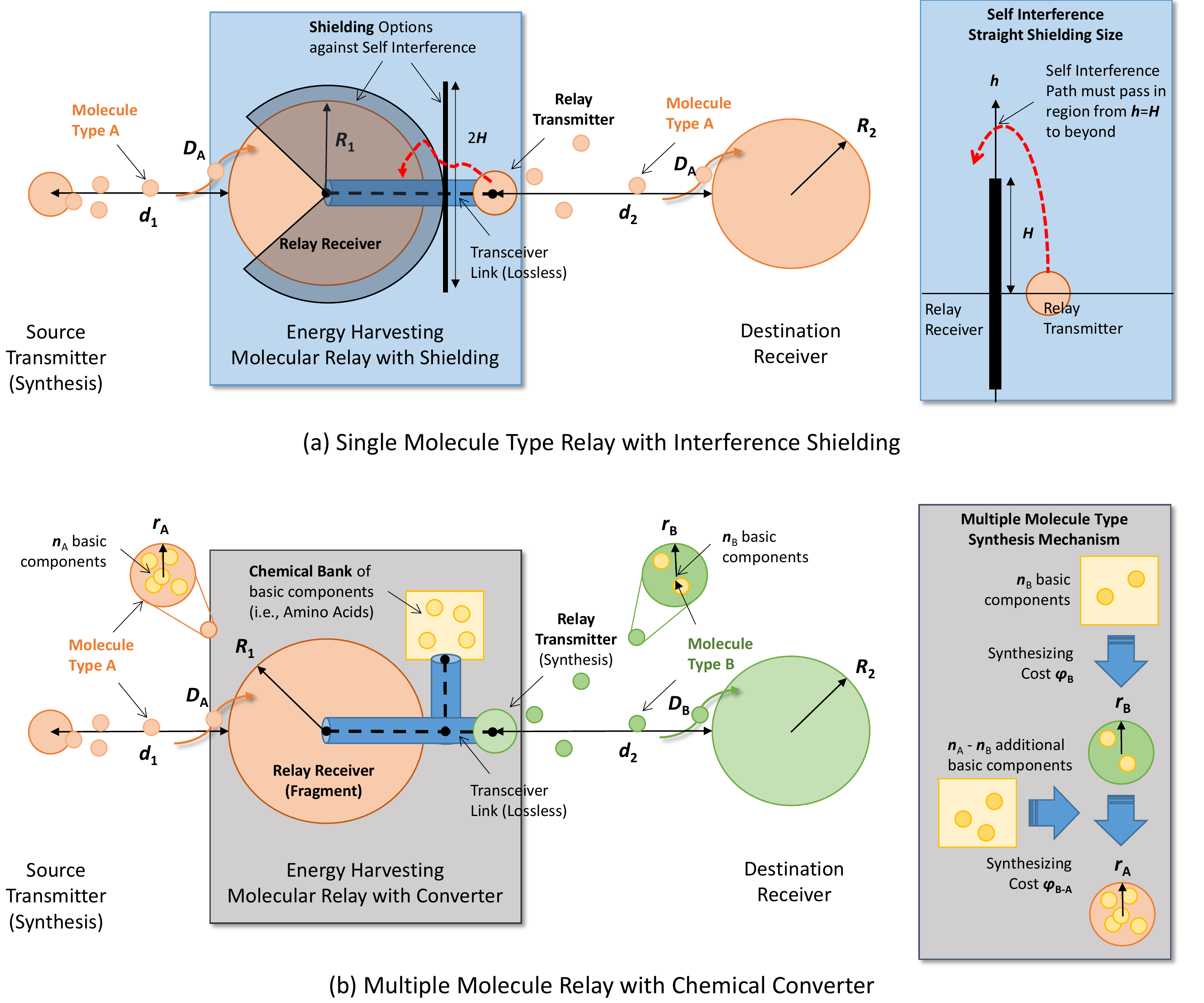}
	\caption{Illustration of energy harvesting MCvD nano-relay systems: (a) single molecule type with self-interference shielding options, and (b) multiple molecule types with chemical converter.}
	\label{System_Relay}
\end{figure*}

Comparing MCvD with RF propagation at the macro-scale to draw similar levels of performance (see Fig.~\ref{Distance_Decay}), the received RF power is $\propto d^{-\alpha}$, where $\alpha$ typically varies from 2 to 4. On the other hand, the received molecules from MCvD can asymptotically be $\propto d^{-1}$, and at best independent of $d$ in ${\mbox{1-D}}$ space, provided the receiver is willing to wait for a long time $t \!\to\! \infty$. Yet, the long waiting time is not as ridiculous as it may appear for two reasons. First, the rate of diffusion is in reality accelerated by ambient air flow (i.e. convection currents) or shortened to a few mili-seconds at the nano-scale. Second, when one transmits a continuous stream of symbols, the power emitted for the first symbol will be recovered by the $N$-th symbol's time (when $N$ is large). Hence, there are no incurred delays to the power or energy recovery in MCvD are incurred, providing a long stream of symbols are transmitted.

\section{Designing Nano-Relay SMIET}

In order to utilize the energy-harvesting concept, one needs to have a system where the absorbed molecules are reused for new transmissions. We propose two types of SMIET nano-relays that can achieve low energy nano-scale communications. We propose a 2-hop energy-harvesting relay system, where, as illustrated in Fig.~\ref{System_Relay}, the relay is capable not only of demodulating the information transmitted by the source at a distance $d_1$ away, but also of harvesting energy by collecting the absorbed molecules. With the absorbed molecules at the relay, the relay transmitter then re-emits the received molecules to the intended destination at  distance $d_2$ away. 

\subsection{Single Molecule Type: Self Interference}
We first consider the case where only one molecule type is used on both links of the 2-hop relay. That is to say, every receiver can absorb the molecule transmitted at the source and retransmitted at the relay. Hence, in this scheme, the transmitter causes interference at the receiver in later symbol slots due to longer propagation time. As illustrated in Fig.~\ref{System_Relay}a, this kind of system will also induce self-interference at the relay's receiver, as the relay transmitter's emission will likely be immediately absorbed by the relay's own receiver. Hence, this type of system requires self-interference shielding at the relay between the receiver and transmitter. In particular, we propose the following: a) the relay transmitter needs to be at a certain distance away from the receiver, and b) be separated by a physical shield in order to avoid any emitted molecules being immediately absorbed by the relay receiver. The study considers two shielding options are considered: (i) a spherical shield that partially protects the relay receiver, so that molecules can enter only from the source transmitter side, giving the relay receiver directionality \cite{Felicetti15}, and (ii) a straight shield that separates the relay receiver and transmitter, similar to a knife-edge channel \cite{Guo15TMBMC}. The latter shield design can be relatively easily analyzed, where any potential self-interference at the relay will need to transverse via the region of $h=H$ to $+\infty$ (see Fig.~\ref{System_Relay}a right panel diagram). This transition probability can be found to be: $\int_{H}^{+\infty} 2f(t,h)\dif h$, where $2f$ is the hitting distribution applied to a hemisphere with the shield acting as a perfectly reflecting boundary. The resulting self-interference probability is therefore $\propto \text{erfc}(H)$, which is intuitively minimized for a large value of $H$. The specific design, especially in a nano-machine sense will depend on the channel diffusivity $D$ and other parameters beyond the scope of this article. 

One example of such a design is the binary concentration shift keying (CSK) with molecule type A~\cite{WCMag16}. In the first source-to-relay (SR) link, the source emits $N_{\text{Tx-S}}$ molecules to transmit bit-1 ($a=1$), and emits 0 molecules to transmit a bit-0 ($a=0$). The relay simultaneously receives information and energy from the SR link. The expected number of adsorbed molecules ($N_{\text{Rx-SR},a}$) at the relay during a bit interval $N_{\text{Rx-SR}}[K]$, which  contributes by the absorbed molecules due to the previous $K-1$  bits transmission at source. The molecules retrieved from the SR link can be decoded and used directly as molecules to carry data to the destination via the relay-to-destination (RD) link. The total transmitted molecules from the relay is under the power budget of $N_{\text{Rx-SR}}[K]$. Therefore, the RD link cannot reliably encode by using the SR energy-harvesting process alone. Appropriate safeguards need to be designed to improve transmission reliability, such as introducing latency to either only transmit when there is sufficient energy to ensure reliability, or when the 2-hop channel is reliable enough \cite{Huang13}. We discuss these techniques in greater detail in Section IV and leave detailed research in this area for the community.

\subsection{Multiple Molecule Types: Fragmentation and Synthesis}
We now consider the case where two molecule types are used at the two links of the relay. In general, this can be extrapolated to multiple hops with multiple molecule types. Utilizing more than one molecule type is attractive for two reasons. First of all, as mentioned in the previous section, the motivation is to avoid self-interference at the relay. By using a different molecule type (i.e., molecule type B), we can avoid the necessity of a shield and significant separation distance between the relay transmitter and receiver. Secondly, as this section will explain, by changing the molecular composition, the relay is able to control the rate of diffusion through the relationship between molecule size and the rate of diffusion, which has been recently explored in \cite{Furubayashi16}. 

As illustrated in Fig.~\ref{System_Relay}b, this kind of system requires all the molecule types to be constructable from basic chemical components such as amino acids (see GABA example in Section I). Therefore, a chemical bank with the building components is required at the relay, such that molecules of type A can be transformed into type B, either by adding new components or removing existing ones. The cost of adding an amino acid to the chain is independent of the type of the amino acid. Therefore, the difference between type A and type B molecules determines the energy efficiency for synthesizing type B molecules from the harvested molecules at the relay node. In this particular example, molecule type A is composed of $n_{A}$ amino acids and molecule type B is composed of $n_{B} \neq n_{A}$ amino acids. As mentioned in Section II, the cost to producing a single molecule that comprises of $n$ amino acids is \cite{Kuran10}: $(n\!-\!1)\phi$, where $\phi$ is the synthesising cost and is given as $\phi\!=\!\SI{202.88}{\zepto\joule}$.

\subsubsection{SR Link}
\begin{figure}[t]
	\centering
	\includegraphics[width=1.00\columnwidth]{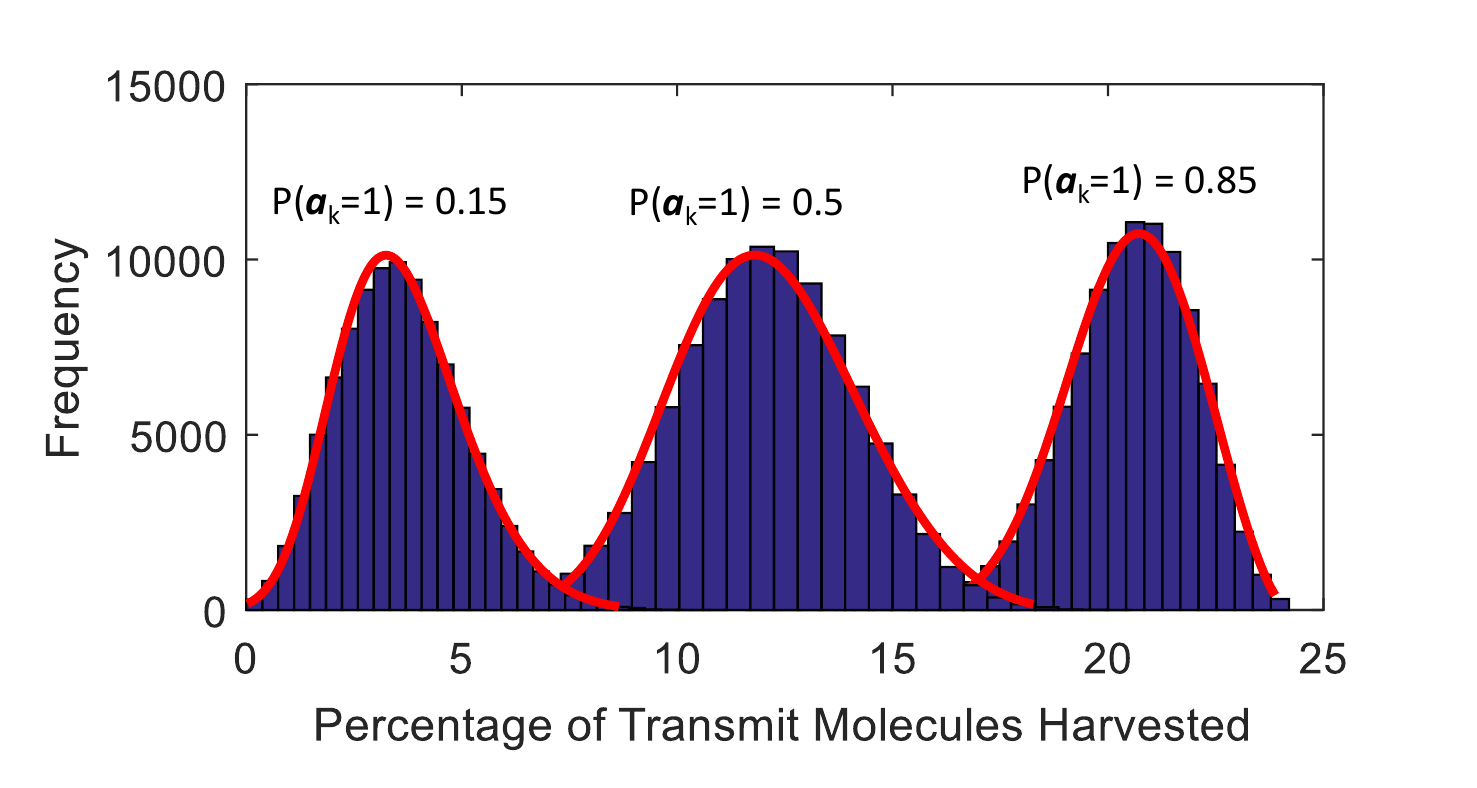}
	\caption{Histogram and distribution fit plots of harvested number of molecules at the relay receiver with varying line-coding bit-1 probabilities $\mathbf{P}(a_{k}=1)$. The distribution of molecules harvested can be modelled as a GEV and a distribution fit is shown with modeling parameters $D= \SI{1e-5}{\centi\metre^2/\second}$, $d=1$mm, and $R=0.2$mm. In all cases, $K=50$ bits were considered with an 1000 iterations.}
	\label{Harvested_Histogram}
\end{figure}

In the first SR link, molecule type A is used to carry data and, as shown in the previous section, the expected number of absorbed molecules ($N_{\text{Rx-SR},a}$) at the relay during the $k$-th bit can generally be expressed as: $N_{\text{Rx-SR}}[K] = N_{\text{Tx-S}}\sum_{k=0}^{K-1} a_{k}[F_{c}((k+1)T)-F_{c}(kT)]$ for $K-1$ previous bits absorbed. 
If all bits are 1 (i.e., $a_{k}=1$ for all $k$ for some line-code), the energy harvested during bit interval $T$ is equal to the total energy available to be harvested: $N_{\text{Rx-SR}}[K] \!\to\! \frac{R_1}{R_1+d_1}  \text{erfc}(\frac{d}{\sqrt{4DKT}})$ which asymptotically converges to $\frac{R_1}{R_1+d_1}$. Figure~\ref{Harvested_Histogram} shows histogram plots of harvested number of molecules at the relay receiver with varying line-coding bit-1 probabilities: $\mathbf{P}(a_{k}=1)=[0.15, 0.5, 0.85]$. The results show that the distribution of the number of molecules harvested can be described by a Generalized extreme value distribution (GEV) with varying parameters, i.e., $N_{\text{Rx-SR}} \sim \text{GEV}(\mu,\sigma,\zeta)$. In general, the energy harvesting potential in the SR link is quite high (approximately 12-25\% for $\mathbf{P}(a_{k}=1)>0.5$) at a distance of 1mm. As the technology for nano-machines shrink and the inter-distance between machines also reduce, the potential to harvest energy improves linearly (see Eq.\eqref{Power_Equation}). 

\subsubsection{RD Link}
In the second RD link, molecule type B is used. Here, we have an opportunity to design type B such that it is sufficiently dissimilar to type A to avoid chemical interference, and to maximize the communication link performance. It is well known that in the limit of low Reynolds number (laminar flow), the mass diffusivity parameter $D$ is inversely proportional to the dimension of the molecule through the \textit{Stokes-Einstein} relation: $D \propto r$, where $r$ is the radius of the molecule which is related to the number of amino acids required for each molecule \cite{Furubayashi16}. By controlling the number of amino acids in molecule type B in the RD link, the relay can control the following:
\begin{itemize}
  \item The total number of molecules (type B) available for transmission, $N_{\text{Tx-R}}[K] = N_{\text{Rx-SR}} [K]\frac{n_{A}}{n_{B}}$. Reducing the number of amino acids in type B will allow more molecules to be available for transmission.
  \item The rate of diffusion $D \propto n_{B}^{-1/3}$ (see \textit{Stokes-Einstein} relation). Increasing the diffusivity of type B molecules will yield a higher peak signal response $\propto D$.
\end{itemize}
Therefore, with multiple molecule types permitted, there is a real freedom to dynamically adjust the chemical composition to suit communication needs. Three scenarios exist for modifying the size of the molecule in the RD link (assuming same constituent amino acid sub-chains):
\begin{itemize}
 \item In the case of $n_{A}=n_{B}$, there is no change in the size of the molecule and no synthesis power cost $\phi$ at the relay. Due to the fact that some molecules will be lost in the SR link, the performance in the RD link will be weaker as in the scenario without any relay assistance. 
 \item In the case of $n_{A}>n_{B}$, the type B molecules are smaller, of higher concentration, and diffuse faster, resulting in a stronger RD link performance than the aforementioned $n_{A}=n_{B}$ scenario. The cost of synthesis is potentially high, as all the amino-acids from type A molecules may need to be disassembled and re-synthesized into type B molecules, incurring a potentially high $\phi$ power cost.
 \item In the case of $n_{A}<n_{B}$, the type B molecules are larger, fewer (of lower concentration), and diffuse slower, resulting in a weaker RD link performance than the aforementioned $n_{A}=n_{B}$ and $n_{A}>n_{B}$ scenarios. However, the propagation loss in the SR link is smaller than that in the RD link, due to $D_A>D_B$, which enable higher energy harvested from the SR link.
\end{itemize}
We leave to future research a detailed study on how to optimize the SR and RD link regarding the number of amino acids per molecule type, the relative distance, and  the effect it has on the rate of diffusion and communication performance is left for future research. Nonetheless, it is worth mentioning that due to the long tail distribution of the diffusion channel response, there is a complex trade-off between the energy-harvesting potential and the high inter-symbol-interference (ISI) in MCvD systems. Certainly, existing research on optimizing resource allocation in relays can be adjusted to MCvD SMIET systems, and this is discussed in greater detail in Section IV.

\section{Learning from Existing SWIPT Techniques}

\subsection{Resource Optimization}

In RF SWIPT, the relay node receives information and harvests energy from the source's transmissions. If, due to hardware limitations, the relaying node cannot perform simultaneously information decoding and energy harvesting, then it is worthwhile dividing current SWIPT research into time-switching (TS) and power-splitting (PS) schemes. Furthermore, due to the unreliable nature of energy harvesting from the MCvD (see Fig.~\ref{Harvested_Histogram} for the GEV distribution of energy arrival in Section III), scheduling becomes very important, as the energy may not arrive with the same quantity or at all in any given time frame. To achieve efficient scheduling, modeling the arrival distribution either as a Markov model or a probabilistic model is useful for the construction of algorithms such as advance-before-scheduling. Researchers also need to consider the finite level of energy or molecular storage in a nano-machine, and the impact it has on SWIPT as well as on processing power expenditure needs to be considered. Existing research in relay based SWIPT has largely focused on the optimization of the tune parameters for TS and PS schemes as a function of the channel state information (CSI). The availability of CSI allows for either online or offline computation of optimal scheduling through complex matrix manipulations. However, CSI (even statistical) is notoriously difficult to estimate in MCvD channels due to the unpredictable and rapid changes in the fluid conditions, as well as the movement of the nano-machines themselves. 

In terms of algorithm complexity, solutions that involve matrix manipulations and multiple directional water-filling computes are complex and not suitable in the context of simple low-energy nano-machines. Therefore, to adapt SWIPT optimally, researchers need to devise non-coherent and low-complexity signal processing and scheduling algorithms. 
\begin{figure}[t]
	\centering
	\includegraphics[width=1.00\columnwidth]{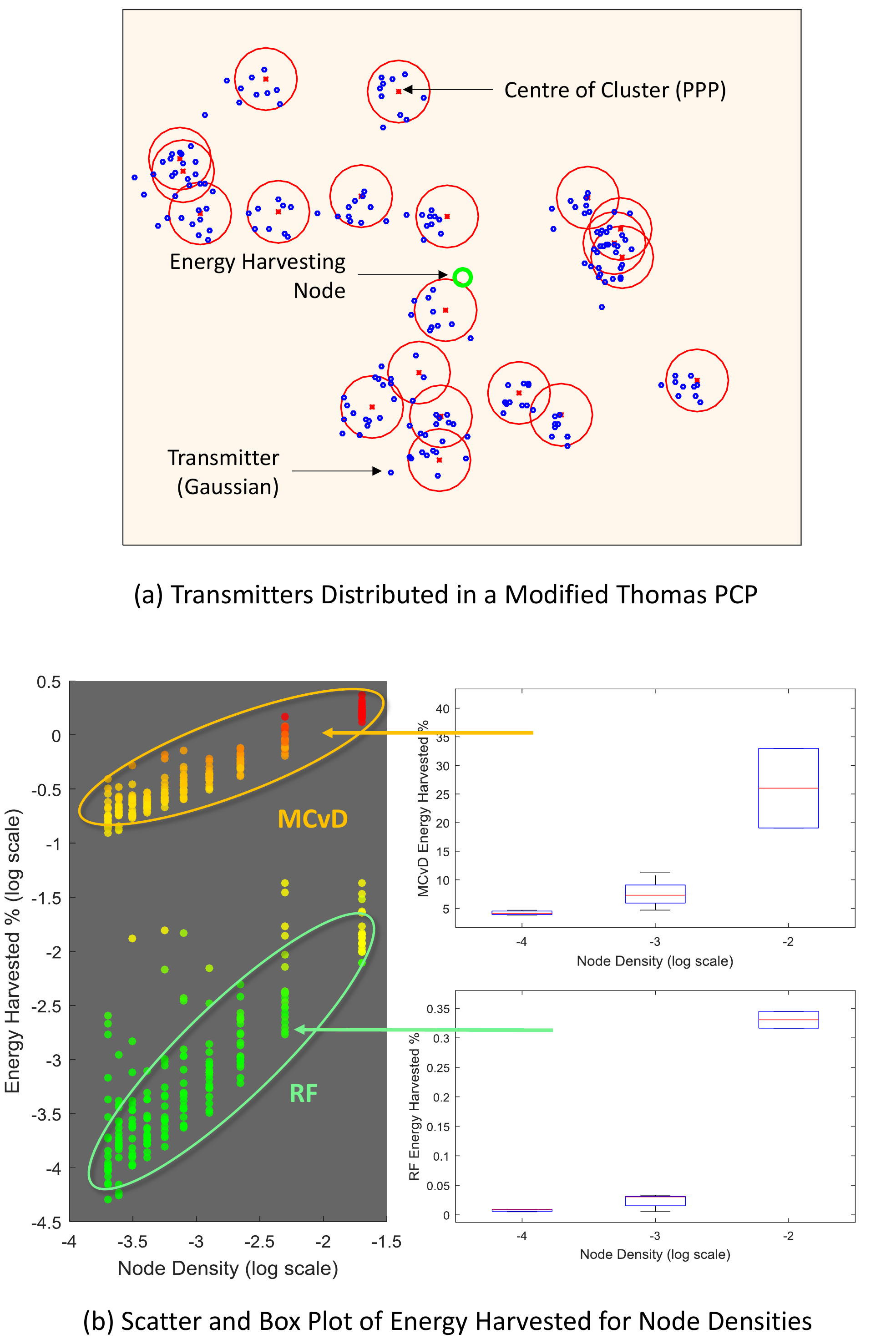}
	\caption{Crowd energy harvesting from a formation of nodes distributed according to a Thomas PCP. Subplot (a) shows an instant snap-shot of the PCP formation of nodes with an energy harvesting receiver at the centre. Subplot (b) shows a scatter and box plot of the percentage of energy harvested for MCvD and RF transmissions as a function of node density (per m$^{2}$). Modeling parameters: a variable modeling area radius 0.1-1km, $20$ clusters each with 10 nodes 2D Gaussian distributed with s.d. 30m, mass diffusivity $D=\SI{79.5}{{\micro\metre}^{2}/s}$, pathloss exponent $\alpha=2$, transmit power $P_{\text{Tx}}=1$W, transmit molecule $N_{\text{Tx}}=1$, and receiver radius of $R=\SI{1}{\metre}$.}
	\label{Crowd_Harvest}
\end{figure}

\subsection{Crowd Energy Harvesting}

Crowd energy harvesting is a popular concept proposed for battery-powered wireless nodes (e.g., low-power sensors) to harvest energy from high-power transmissions from TV broadcasts and cellular base stations. Despite the growing density of RF transmissions across multiple spectrum bands, the amount of energy available to harvest is dominated by the closest high-power link. The rapid loss in RF energy due to transmission distance limits the potential for crowd harvesting, and unless all the transmitters are spaced equidistant to the receiver, crowd harvesting energy from $N$ transmitters is not significantly superior to the receiving energy from the nearest transmitter. For MCvD systems, as mentioned in Section II, the energy of molecules do not obey the propagation laws of waves. Instead of experiencing a hostile $\propto d^{-\alpha}$ rate of energy decay, molecular numbers (or energy) decays $\propto d^{-1}$. Therefore, the potential to harvest energy from a field of transmitters is far greater for MCvD than for RF communications. If one assumes that the molecular transmitters are randomly and uniformly distributed according to a Poisson Point Process (PPP) or Poisson Cluster Process (PCP), one can find the expected molecular energy by leveraging existing stochastic geometry techniques \cite{Akbar16}. This typically involves understanding the general distance distribution $f_{D}(d,n)$ from the energy harvesting node to the $n$-th nearest transmitter node.

Figure~\ref{Crowd_Harvest} shows a simulation of crowd harvesting energy from a formation of nodes distributed according to a modified Thomas PCP. Subplot (a) shows an instant snap-shot of the PCP formation of nodes with an energy harvesting receiver at the centre. Subplot (b) shows a scatter and box plot of the percentage of energy harvested for MCvD and RF transmissions. The results show that RF energy harvesting is far more sensitive to the density of transmitter nodes than MCvD energy harvesting. The random walk nature of molecular propagation means that the distance distribution (i.e., $f_{D}(d,n)$) is not a key consideration in crowd harvesting, whereas it is for RF systems. This demonstrates the potential for crowd harvesting in molecular systems, which can achieve 2-5dB improvement in harvesting efficiency compared to RF systems in a similar setting, with the highest relative gain at low node densities.

\section{Conclusions and Future Work}

The performance of communication systems is fundamentally limited by the loss of energy through propagation and circuit inefficiencies. In this article, we have shown that it is possible to achieve ultra-low energy communications at the nano-scale. We show that while the energy of waves will inevitably decay as a function of transmission distance and time, the energy in molecules does not. In fact, over time, the molecular receiver has an opportunity to recover some, if not all of the molecular energy transmitted. Inspired by the GABA metabolism system, which fragments and reassembles molecules, we design two nano-relay systems that can achieve extremely high energy harvesting efficiency (12-25\%) in realistic nano-scale conditions. We further examine the potential of crowd harvesting energy from a swarm of nano-machines and demonstrate that molecular communications is significantly less sensitive to the spatial distribution of nano-machines and can achieve 2-5dB improvement in harvesting efficiency. 


In terms of algorithms, challenges remain in applying existing radio-frequency energy harvesting solutions to molecular communications, namely the lack of channel state information and the low complexity requirement of any algorithms that operate in nano-machines with limited computational ability. What is more interesting is the need to build functional energy harvesting systems that involves a reservoir of molecules that can be used for fresh transmissions and replenished from received and decoded molecular signals. Inspired by biological processes, researchers can chemically manipulate the chemicals in a reservoir to potentially improve the energy efficiency of information relaying, something that is impossible for radio frequency communications. What currently remains beyond current engineering capabilities is the ability to build such biological functionalists into realistic systems remains beyond current engineering capabilities. Nonetheless, the preliminary results in this article indicate that the potential for simultaneous molecular information and energy transfer (SMIET) is immense and further research is needed to transform theory to reality.

\bibliographystyle{IEEEtran}
\bibliography{IEEEabrv,CM_Ref_new}

\end{document}